\documentclass[a4paper]{article}

\usepackage{icrc2013}

\title{Search for gamma-ray induced showers from the lateral distribution of electrons in EAS}

\shorttitle{Search for $\gamma$-ray induced showers}

\authors{
R. K. Dey$^{{1},{2}}$,
A. Bhadra$^{2}$
}

\afiliations{
$^1$ Department of Physics, University of North Bengal, Siliguri, WB 734013 India \\
$^2$ High Energy and Cosmic Ray Research Centre, University of North Bengal,
Siliguri, WB 734013 India \\
}

\email{rkdey2007phy@rediffmail.com}

\abstract{Distinguishing $\gamma$-ray and hadron initiated extensive air showers (EAS) based on lateral distribution of electrons has been studied by detailed Monte Carlo (MC) simulations. The possibility of using the local age parameter (LAP) of EAS for the gamma-hadron separation has been explored. It is found that separating $\gamma$-ray and hadron induced EAS on the basis of LAP can be useful for surface detector experiments those have no reliable muon measurement facilities.}


\begin{document}
\maketitle

\section{Introduction}
The important features of the primary energy spectrum vis-a-vis the chemical composition of the cosmic rays (CRs) at ultra high energy (UHE) and extremely high energy can be understood from a complete study of CR air-shower. The direct measurements of primary CR flux nearly $100$ TeV and above are impractical because of very low and sharply falling flux, but  has to be inferred from observations of extensive air showers.

To extract information about primary CRs from the measurements of various experiments still require detailed MC simulations of the shower development as a basis of the data analysis and interpretation. The MC simulations consider the evolution of EAS in the atmosphere initiated by different energetic particles.

In the field of CR air shower physics, the discrimination of $\gamma$-ray induced air showers from hadron-induced air showers is a challenging problem that still needs more attention \cite{bib:gaisser}. The high energy $\gamma$-rays are considered to be the by-product of hadronic components of primary CRs: a fraction of accelerated hadronic cosmic rays are likely to produce $\gamma$-rays through $\Delta$-resonances,  bremsstrahlung, inverse Compton etc. processes at or very close to the source site by interacting with the ambient matter \cite{bib:bhadra}. The $\gamma$-ray detection suffers from the huge background constituted by ordinary CRs (hadrons) in the GeV-TeV energy region. To observe astrophysical sources (emitting undeflected $\gamma$-rays and assumed as point-like objects) and to study the anisotropy properties of primary CRs, one has to eliminate isotropically distributed CRs.

Usually poor muon content is considered as the signature of $\gamma$-ray initiated EAS. In order to select a $\gamma$-ray shower based on such a criterion, an EAS array needs to be equipped with muon detectors covering very large area which is economically very challenging and such a facility is rarely available. So one has to look for some other primary mass sensitive observables based on which $\gamma$-ray initiated EASs can be separated out without the need of large area muon detector. The lateral shower age, which is essentially the slope of the lateral density distribution of electrons in EAS reflects the developmental stage of EAS and hence it can be used as distinguishing parameter. 

The selection of $\gamma$-ray initiated showers is attempted here by employing two different approaches - the Method I and the Method II. In the former approach, we have taken single (r-independent) lateral age parameter i.e. $s_{\bot}$ as the $\gamma$-ray separation parameter. Experimentally it is observed that the NKG function with a single lateral age is insufficient to describe the lateral density distribution of EAS electrons properly at all distances, which implies that the lateral age changes with the radial distance. Subsequently, the notion of local lateral shower age parameter (LAP) was introduced \cite{bib:capdevielle} which is in essence the lateral age at a point. Since experimental electron density data in EAS may fluctuate considerably at a particular radial distance, instead of taking LAP at any particular point we take an average LAP between $50$ m (minimum) and $300$ m (maximum) and in Method II, we employed this average LAP ($s_{local}$) to select $\gamma$-ray showers. 

The present study has been performed mainly at the geographical location of ARGO-YBJ \cite{bib:aloisio} (latitude $30.11^{o}$ N, longitude $90.53^{o}$ E, $4300$ m a.s.l.). This is because the experiment offers a full coverage array and hence can measure radial density distribution of electrons in EAS with great accuracy, which in turn provides an opportunity to estimate LAP accurately. However, ARGO-YBJ has not yet studied the radial variation of the lateral shower age while a few other experiments, such as Akeno \cite{bib:nagano} and NBU \cite{bib:sanyal}, successfully tested the predicted radial variation of LAP. We simulated a few events at the geographical location of Akeno and compared with the observations to demonstrate again the importance of considering LAP instead of single shower age. To ensure that the conclusion of the present work is robust, we consider two high energy interaction models, QGSJet 01 v.1c \cite{bib:kalmykov} and EPOS - 1.99 \cite{bib:werner} and found that the present findings do not have any strong dependence on the choice of interaction model. 

\section{Shower age parameters from the cascade theory} 
Nishimura and Kamata \cite{bib:lipari} solved numerically the 3-dimensional shower equations in {\it Approximation B} (electrons suffer constant amount of collision loss in a radiation length which is equal to the critical energy) to obtain the lateral distribution of electrons propagating in a medium of constant density. The results obtained on lateral density distribution of cascade particles by Nishimura and Kamata can be approximated by the well known Nishimura-Kamata-Greisen (NKG) structure function proposed by Greisen \cite{bib:greisen}, given by

\begin{equation}
f(r)=C(s_{\bot})(r/r_{m})^{s_{\bot}-2}(1+r/r_{m})^{s_{\bot}-4.5} \;,
\end{equation}

where the normalization factor C($s_{\bot}$). 

The normalization of $f(r)$ implies for the electron density $\rho(r)= N_{e} f(r)$. But such a widely used relation does not hold if $s_{\bot}$ varies with $r$, as noted in Akeno \cite{bib:nagano}, NBU \cite{bib:sanyal} and some other observations \cite{bib:capdevielle}. Consequent upon, the lateral shower age is found to vary with radial distance experimentally. To handle the situation a method was developed by Capdevielle {\it et al} introducing the notion of {\it local age parameter} (LAP) \cite{bib:capdevielle}. From two neighboring points, $i$ and $j$, we can give a LAP for any distribution $f(x)$ (where$~x={r\over r_{m}}$) which characterizes the best fit by a NKG-type function in [$x_{i},x_{j}$]~:

\begin{equation}
s_{local}(i,j) = {{\ln(F_{ij} X_{ij}^{2} Y_{ij}^{4.5})} \over {\ln(X_{ij} Y_{ij})}}
\end{equation}

Here, $F_{ij}$ = {{$f(r_{i}$)}/{$f(r_{j}$)}}, $X_{ij}$=$r_{i}$/$r_{j}$, and $Y_{ij}$=($x_{i}$+1)/($x_{j}$+1). More generally, if $r_{i} \rightarrow r_{j}$, this suggests  the definition of the LAP $s_{local}(x)$ (or $s_{local}(r)$) at each point~:

\begin{equation}
s_{local}(x) = {1 \over {2x+1}} \left( (x+1) {{\partial{\ln f}} \over {\partial{\ln x}}} + 6.5x + 2 \right)
\end{equation}

Function $f_{NKG}(r)$ with $s$=$s_{local}(r)$ can be used to fit $f(r)$ in the neighborhood of $r$.

The identification $s_{local}(r)\equiv s_{local}(i,j)$ for $r=\frac{r_{i} +r_{j}}{2}$ remains valid for the experimental distributions (taking $F_{ij}~ =~\rho(r_{i})/\rho(r_{j}$)) as far as they are approximated by monotonic decreasing functions versus distance. 

The behavior of LAP on experimental lateral distributions was found in accordance with the prediction \cite{bib:capdevielle et al} which was reaffirmed by the Akeno observations \cite{bib:nagano}. The stated method was validated by the rapporteurs of the ICRC from 1981 to 1985 \cite{bib:tonwar}. 

\section{Simulation of EAS}
The EAS events are simulated by coupling the high energy (above 80 GeV/n) hadronic interaction models QGSJet 01 version 1c \cite{bib:kalmykov} and  EPOS 1.99 \cite{bib:werner}, and the low energy (below 80  GeV/n) hadronic interaction model GHEISHA (version 2002d) \cite{bib:fesefeldt} in the framework of the CORSIKA Monte Carlo program version  6.970 \cite{bib:heck}. For the electromagnetic part the EGS4 \cite{bib:nelson} program library has been used. 

The MC simulation data library consists of considerable amounts of EAS events each for p, Fe and $\gamma$-ray at the ARGO-YBJ and Akeno levels. A mixed sample has been prepared from the generated showers taking $37 \%$ p, $37 \%$ Fe, $26 \%$ $\gamma$-ray events for better understanding of EAS observational results. 

\section{Estimation of lateral shower age} 

The simulated electron density data have been analyzed in two different methods to obtain shower age parameters $s_{\bot}$ and $s_{local}(r)$. First, following the traditional approach we estimate $s_{\bot}$ by fitting the density data with the NKG structure function. The error in estimating $s_{\bot}$ in the $N_{e}$ range $1.5\times 10^{4}-4.25\times 10^{5}$ has been found as $\pm 0.03$.

Secondly, exploiting Eq.(2) we directly estimated LAP for each individual event. In this analysis, the error of the LAP for EAS with the primary energy in the PeV range remains within $0.05$ for $10$ $m < r < 250$ m, whereas for $r < 10$ m or when $r > 300$ m the error of the LAP is found to be higher, about $0.1$.

\begin{figure}[t]
  \centering
  \includegraphics[width=0.4\textwidth,clip]{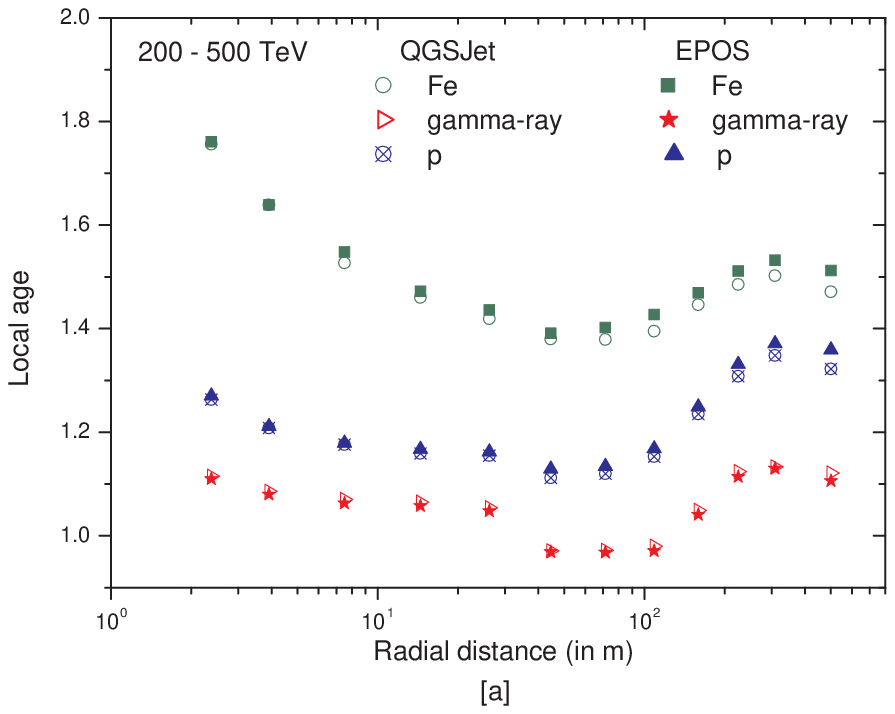} 
\includegraphics[width=0.4\textwidth,clip]{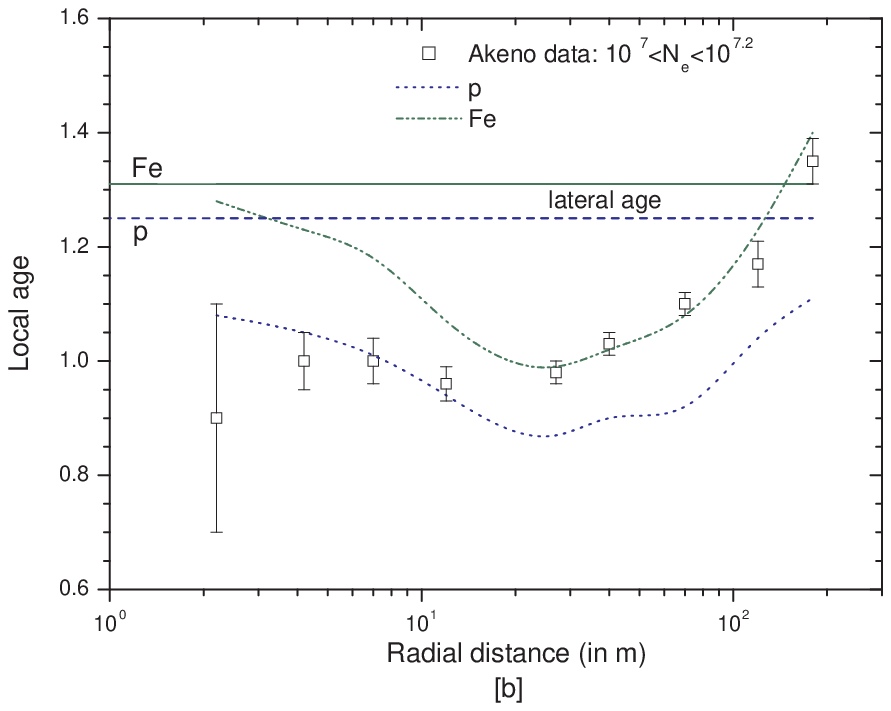}    
\caption{ [a] The radial variation of LAP obtained at the ARGO-YBJ altitude for p, Fe and $\gamma$-ray with hadronic interaction models QGSJet and EPOS. [b] Same as figure 1a but at the Akeno level with QGSJet model, compared with experimental data. The solid and dashed lines parallel to the x-axis indicate $s_{\bot}$ for Fe and p.} 
 \label{fig1_ab}
 \end{figure}

It has been shown recently that LAP initially decreases with radial distance, reaches a minimum around $50$ m, then starts increasing with radial distance, attains a local maximum around $300$ m and decreases again thereafter \cite{bib:dey}. We observed such a characteristic variation of LAP with radial distance at ARGO-YBJ location as shown in the top figure \ref{fig1_ab} for proton, iron and $\gamma$-ray primaries for both QGSJet and EPOS models. 

As already mentioned, Akeno group studied the radial variation of LAP experimentally \cite{bib:nagano}. We compared our simulation results with Akeno observations in the bottom figure \ref{fig1_ab} for proton and Fe primaries. The single lateral shower age for proton and iron primaries are also given (solid and dashed lines  parallel to the x-axis) in the figure for comparison. 

 Since air shower measurements are subjected to large fluctuations, instead of LAP at a particular radial distance we consider for each event a mean LAP ($<s_{local}>$), which is the average of LAPs for several small distance bands ($r_{i},r_{j}$) over the radial distance between $50$ m to $300$ m. For the purpose of averaging, distance bands are taken in constant steps on the logarithmic scale. The radial distance band from $50$ m to $300$ m is chosen because the positions of local minimum and maximum at $50$ m and $300$ m are nearly universal, independently of primary energy \cite{bib:dey}.   

\section{Potentiality of Method II over Method I}
\begin{figure}[t]
\centering
\includegraphics[width=0.4\textwidth,clip]{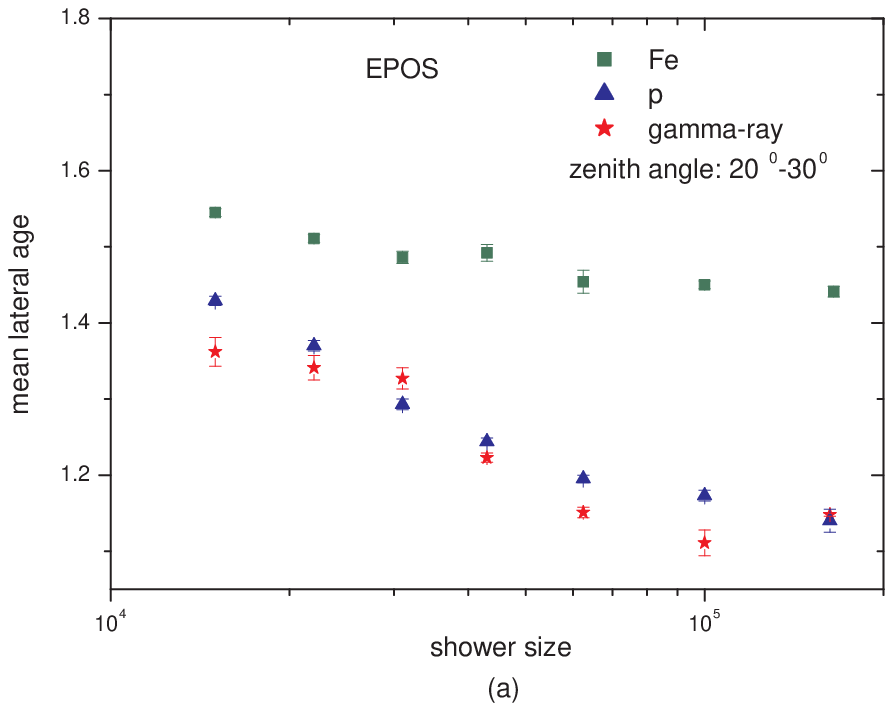}
\includegraphics[width=0.4\textwidth,clip]{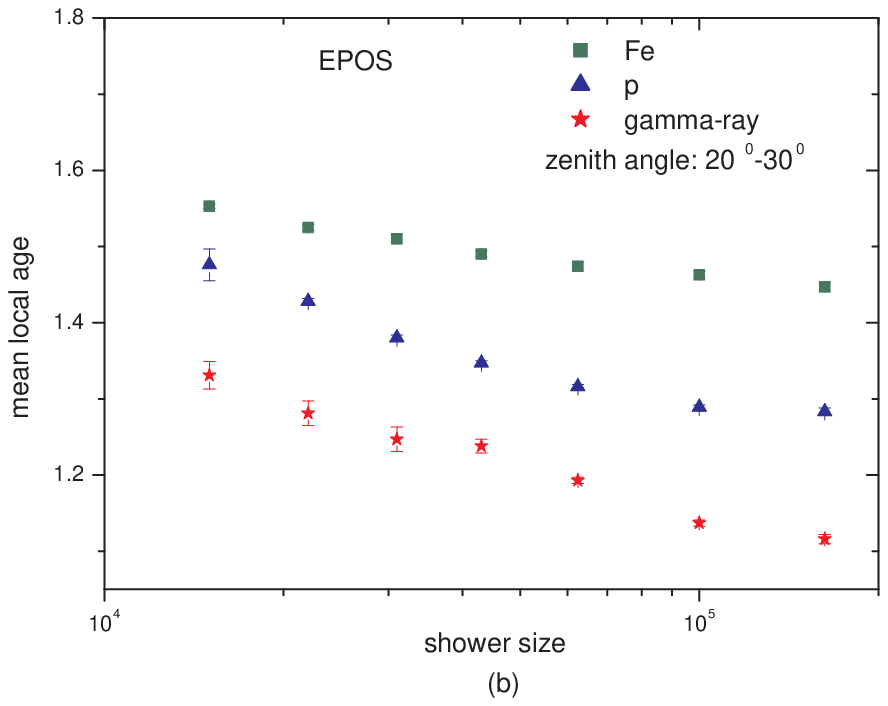} 
\caption{[a] Variation of mean lateral shower age with shower size at ARGO-YBJ level; [b] Variation of mean local age parameter with shower size at ARGO-YBJ level.}
\label{fig2_ab}
\end{figure}

The inadequacy of a single (constant) lateral age parameter to describe the experimental lateral distribution of EAS electrons properly at all distances has been noted in several experimental observations. So when a single constant age is assigned to an EAS event, the discriminating power of that parameter on primary masses somewhat becomes dull; it still can distinguish iron initiated showers from proton induced showers but can't effectively separate out $\gamma$-ray showers from EASs generated by primary protons and this is clear from the comparison between the top and bottom figures in \ref{fig2_ab}. 

\begin{table*}
\begin{center}
\begin{tabular}
{|l|c|c|c|c|c|c|c|c|c|} \hline
     $s_{local}$     &$1.05$     &$1.08$     &$1.10$     &$1.12$     &$1.14$     &$1.15$     &$1.16$     &$1.18$     &$1.20$ \\     
&&&&&&&&& \\ \hline
    
     $\epsilon_{\gamma}$     &$0.146$     &$0.237$     &$0.348$     &$0.510$     &$0.657$     &$0.707$     &$0.778$     &$0.864$     &$0.934$ \\ 
&&&&&&&&& \\ \hline
    
     $\epsilon_{bkg}$     &$0.007$     &$0.022$     &$0.029$     &$0.036$     &$0.058$     &$0.065$     &$0.109$     &$0.145$     &$0.254$ \\ 
&&&&&&&&& \\ \hline

     $Q$     &$1.74$     &$1.61$     &$2.05$     &$2.69$     &$2.73$     &$2.77$     &$2.36$     &$2.27$     &$1.86$ \\
&&&&&&&&& \\ \hline 

\end{tabular}
\caption {The quality factor at various $s_{local}$ cuts using QGSJet model. The primary energy, zenith angle and shower size intervals are $200-500$ TeV, $20^{0}-30^{0}$ and $(1.5 - 4.0)\times 10^{5}$ respectively.} 
\label{table_1}
\end{center}
\end{table*}

\begin{table*}
\begin{center}
\begin{tabular}
{|l|c|c|c|c|c|c|c|} \hline
      Model   & $E$ (TeV)   & $\theta$ (deg.)   & $N_{e}\times 10^{5}$   & {$s_{local}$}   & $\epsilon_{\gamma}$   & $\epsilon_{bkg}$   & $Q$ \\ 
&&&&&&& \\ \hline
    
      EPOS   & $100-200$   & $5-15$   & $0.6-2.0$   & 1.14   & 0.691   & 0.072   & 2.57 \\ 
&&&&&&& \\ \hline

      EPOS   & $200-500$   & $20-30$  & $1.5-4.0$   & 1.15   & 0.734   & 0.048   & 3.35 \\ 
&&&&&&& \\ \hline

      QGSJet   & $200-500$   & $20-30$   & $1.5-4.0$   & 1.15   & 0.707   & 0.065   & 2.77 \\ 
&&&&&&& \\ \hline

\end{tabular}
\caption {The signal selection parameters at optimal conditions.} 
\label{table_2}
\end{center}
\end{table*}

\begin{figure}[t]
\centering
\includegraphics[width=0.4\textwidth,clip]{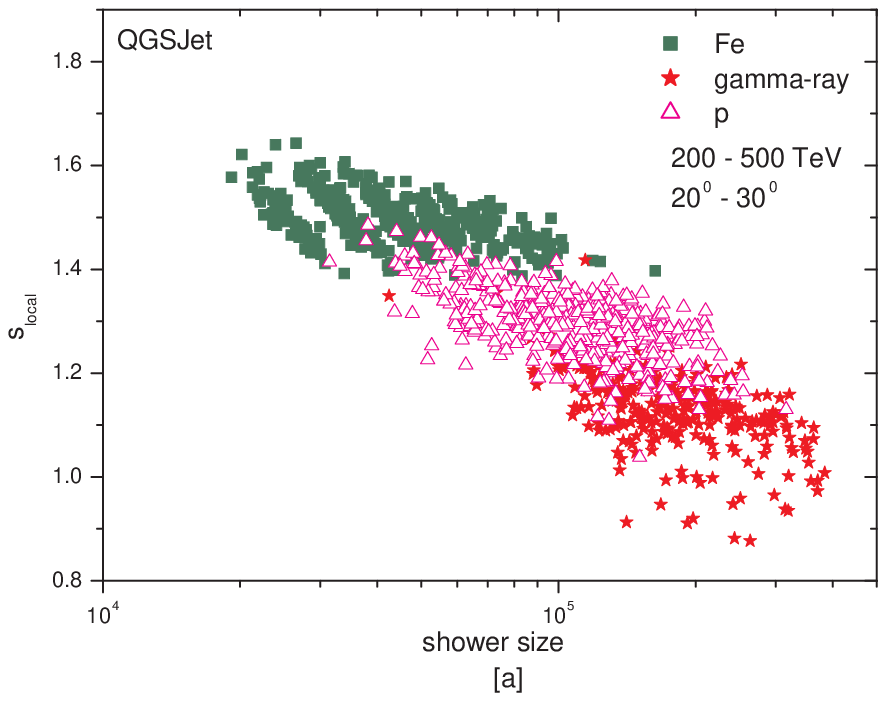} 
\includegraphics[width=0.4\textwidth,clip]{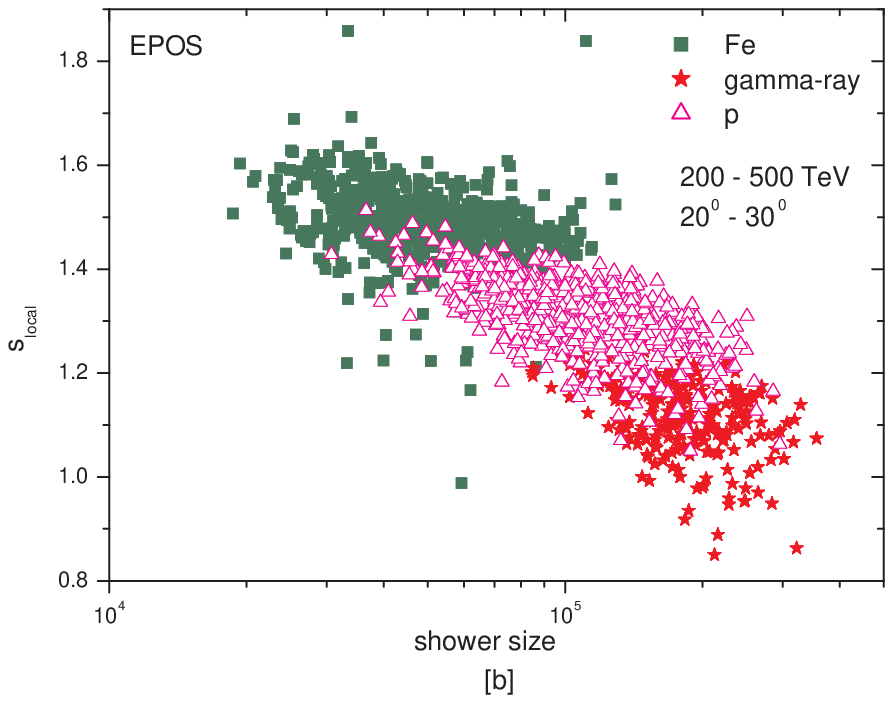}
\caption{Distribution of simulated events in mixture-II in a primary energy range based on $s_{local}$ and $N_{e}$ at ARGO-YBJ.Dependence on interaction models is also shown.}
\label{fig3_ab}
\end{figure}

\section{Selection of ${\gamma}$-ray component using LAP:Method II}
In the primary CR flux, the percentage of $\gamma$-ray flux is very very small, of the order of $0.001 \%$. To separate such a small fraction of $\gamma$-ray component from primary CRs in a real experiment employing method II, it would be nice if we could prepare a mixture with such a small percent of $\gamma$-rays and see whether the $\gamma$-ray events can be extracted out or not. But due to limited statistics we could not do that. Instead, we made three sub-mixtures of type II maintaining the ratio of primaries as $37 \%$ p, $37 \%$ Fe and $26 \%$ $\gamma$-ray for three different combinations of zenith angle, primary energy and shower size range (scatter plots in figures \ref{fig3_ab}). These exercises are done for both the high energy interaction models QGSJet and EPOS. From each sub-mixture, we tried to separate $\gamma$-ray showers out exploiting $s_{local}$ corresponding to different $N_{e}$. 

For a small enough cut value of $s_{local}$ we have found poor acceptance for $\gamma$-ray induced EAS and very good rejection of background, whereas for a large enough value of $s_{local}$ the situation is found to be completely reversed. Cut values of $s_{local}$ lie between the two extremes offered different acceptances and rejections of $\gamma$-ray and background respectively. Similarly a proper cut on $N_{e}$ is also needed for selection of $\gamma$-ray induced EAS and the rejection of background. In astronomical signal selection, optimal cut to selection parameters (maximizing the $\gamma$-ray efficiency and minimizing the background contamination) is usually set by numerical maximization of the quality factor $Q$ defined by 
 
\begin{equation}
Q=\frac{\epsilon_{\gamma}}{\sqrt{\epsilon_{bkg}}}=\frac{\epsilon_{\gamma}}{\sqrt{(1-\xi_{bkg})}}
\end{equation}

where $\epsilon_{\gamma}$ and $\epsilon_{bkg}$ respectively denote the acceptances of $\gamma$-ray and background from the sample using a cut value of $s_{local}$ but $\xi_{bkg}$ stands for the rejection of background. $Q$ essentially quantifies the gain of significance achieved by the separation algorithm. As for an example the quality factor is evaluated over a shower size and zenith angle bins by varying $s_{local}$ which is shown in table (Table \ref{table_1}). In table (Table \ref{table_2}), we have given a chart for quality factors estimated at optimal conditions in three different situations.

\section{Summary \& Conclusions}
 
In this work, we first explored the $r$-independent shower age as the distinguishing parameter. Later, we attempted to separate out $\gamma$-ray initiated EAS on the basis of mean LAP. It is found from the simulation results that $\gamma$-ray induced EASs are younger in terms of the mean LAP and hence this parameter (mean LAP) can effectively separate out $\gamma$-ray induced showers from hadronic EAS unlike the case of single lateral shower age. Lateral distribution of electrons in EAS exhibits universal (primary energy and mass independent) behavior in terms of LAP \cite{bib:dey}. 

An important question is the experimental realization of the adopted technique involving LAP. The uncertainty in estimating LAP is usually large in normal circumstances in comparison to that in lateral shower age as the LAP depends on the logarithmic derivative of the density versus radial distance. These uncertainties should be small for a closely packed air shower array like GRAPES-III at Ooty \cite{bib:gupta} or for a full coverage EAS array like ARGO-YBJ \cite{bib:aloisio}.

\vspace*{0.5cm}
\footnotesize{{\bf Acknowledgment:} {We are thankful to Prof. J. N. Capdevielle for many useful suggestions. RKD thanks the UGC (Govt. of India) for support under Grant. No. 41/1407/2012(SR.) and North Bengal University for providing NBU Research grant. w.e.f. 2012. RKD also gratefully acknowledges the support of DST (Govt. of India) under International Travel Scheme 2013}. }

\end{document}